\documentclass[twocolumn,english]{revtex4}
\usepackage[]{fontenc}
\usepackage[latin1]{inputenc}
\usepackage{graphicx}

\makeatletter

\usepackage{babel}
\makeatother
\begin{document}

\title{Optimized multicanonical simulations: a new proposal based on classical
fluctuation theory}

\author{J. Viana Lopes}

\affiliation{Centro de Física do Porto and Departamento de Física, Faculdade de
Ciências, Universidade do Porto, 4169-007 Porto, Portugal }

\affiliation{Departamento de Física, Instituto Superior de Engenharia, Instituto
Politécnico do Porto, Porto Portugal}

\author{Miguel D. Costa and J.M.B. Lopes dos Santos }

\affiliation{Centro de Física do Porto and Departamento de Física, Faculdade de
Ciências, Universidade do Porto, 4169-007 Porto, Portugal }

\author{and R. Toral}

\affiliation{Instituto Mediterraneo de Estudios Avanzados IMEDEA (CSIC-UIB). Ed.
Mateu Orfila, Campus E-07122 Palma de Mallorca Spain.}

\begin{abstract}
We propose a new recursive procedure to estimate the microcanonical
density of states in multicanonical Monte Carlo simulations which
relies only on measurements of moments of the energy distribution,
avoiding entirely the need for energy histograms. This method yields
directly a piecewise analytical approximation to the microcanonical
inverse temperature, $\beta(E)$, and allows improved control over
the statistics and efficiency of the simulations. We demonstrate its
utility in connection with recently proposed schemes for improving
the efficiency of multicanonical sampling, either with adjustment
of the asymptotic energy distribution or with the replacement of single
spin flip dynamics with collective updates. 
\end{abstract}
\maketitle

\section{Introduction}

Over the past decade, the use of Monte Carlo methods \cite{NewmanBarkema}
has broken the boundaries of statistical physics and has become a
widely used computational tool in fields as diverse as chemistry,
biology and even sociology or finance. 

Despite the enormous success of the well known Metropolis importance
sampling algorithm, its narrow exploration of the phase space and
characteristic convergence difficulties motivated a number of different
approaches. Firstly, the techniques for harvesting useful information
from the statistical data obtained at a given temperature were improved
in order to extrapolate the results in a small temperature range and,
therefore, reduce the number of required independent simulations \cite{FS88,FS89}.
Secondly, cluster update algorithms were proposed \cite{NIE88,WOL89,SW87}
in order to overcome critical slowing down. Finally, the requirement
of constant temperature was lifted, allowing the system to explore
a wider range of the energy spectrum. In Simulated Tempering, the
temperature becomes a dynamical variable, which can change in the
Markov process \cite{MP92}; Parallel Tempering uses several replicas
of the system running at different temperatures and introduces the
swapping of configurations between the various Markov chains \cite{HN96,TvO+96,HTY98}.
The flat histogram approach \cite{BC92b,LEE93-a} replaces the asymptotic
Boltzmann distribution for the asymptotic probability of sampling
a given state of energy $E_{i}$, $p_{i}\propto\exp(\beta E_{i})$,
by $p_{i}\propto1/n(E)$, $n(E)$ being the microcanonical density
of states, thus ensuring that every energy is sampled with equal probability. 

The first major obstacle to this last approach is the obvious difficulty
of accessing the true density of states of a given system; several
clever algorithms have been proposed to this end \cite{BC92b,WL01b}.
Nevertheless, even in cases where the true density of states is known
a priori, recent studies \cite{DTW+04,CLLdS} have shown that very
long equilibration times can remain a serious concern with multicanonical
methods. Furthermore, the number of independent samples is strongly
dependent on energy, making error estimation rather tricky \cite{LLdS}.
This has led some authors (\cite{MPR+01}) to question the reliability
of the results obtained by the Multicanonical Method in spin glass
models at low temperatures. To address these issues, the requirement
of a perfectly flat histogram was also lifted \cite{THT04}, sacrificing
the equal probability of sampling each energy in favor of minimizing
tunneling times. 

In section II, we propose a variation of the algorithm to estimate
the density of states that does not use histograms, instead relying
entirely of measurements of cumulants of the energy distribution at
each stage of the simulation to build a piecewise analytical approximation
to the statistical entropy, $S(E)=\ln\left(n(E)\right)$, and inverse
temperature $\beta(E)=dS(E)/dE$. We find that the time required to
explore the entire energy spectrum scales more favorably with system
size than histogram based methods. The method is as easy to apply
in systems with continuous spectrum as in the discrete case, can be
quite naturally adapted to running a simulation in a chosen energy
range, and accommodates without difficulty tunneling times optimization
schemes and cluster update methods.

In Section III we show that the optimization scheme proposed in \cite{THT04}
can be applied during the process of estimating the density of states,
still avoiding histograms, and with significant efficiency gains.

In Section IV, we demonstrate the usefulness of the analytical approximation
of $\beta(E)$ in generalizing Wolff's cluster algorithm \cite{WOL89}
to multicanonical simulations. This generalization maintains an acceptance
probability still very close to unity, growing large clusters at low
temperatures and small ones at higher temperatures. The optimization
procedure reported in \cite{THT04} is also implemented for this cluster
dynamics. 

This algorithm has already been applied with success to both discrete
(Ising models on regular lattices and Small World networks, Ising
Spin Glasses)\cite{VLP+04} and continuous models (XY and Heisenberg
models, with both short and long-range interactions, namely dipolar
interactions)\cite{Cos01}, but this is its first systematic presentation.

\section{Proposal }

\begin{figure}[htbp]
\begin{centering}\includegraphics[width=0.8\columnwidth]{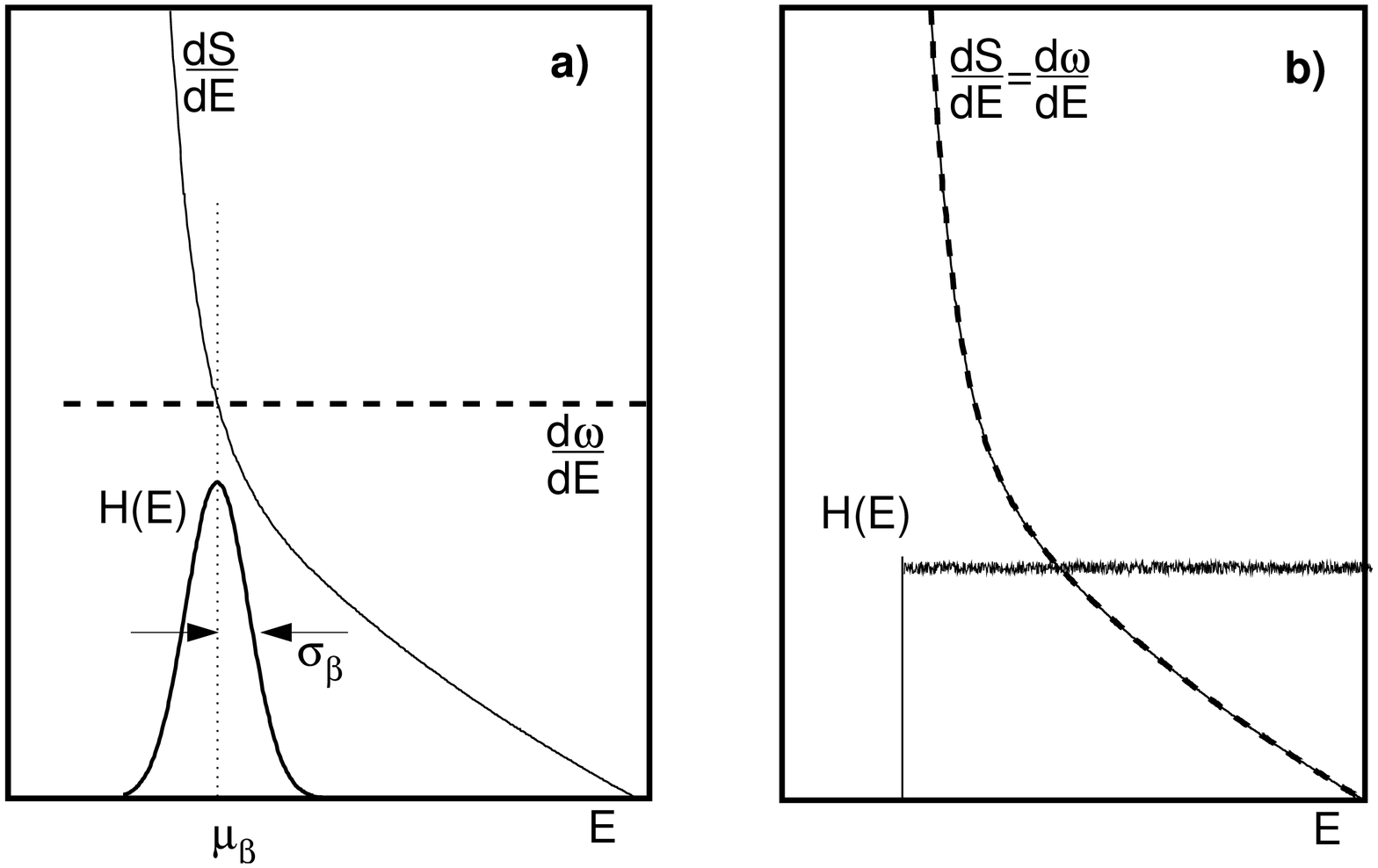}\par\end{centering}

\caption{\label{esquemas}Schematic representation of the true derivative
of the entropy $S(E)$ and two different choices for the derivative
of $\omega(E)$, with the corresponding energy probability distributions.
In panel a), the Metropolis approach, where $\omega(E)=\beta E$ and,
in panel b), the ideal extended ensemble approach with $\omega(E)=S(E)$. }
\end{figure}

The usual way of ensuring an asymptotic distribution $p_{i}\propto\exp\left[-\omega(E_{i})\right]$
is to use a Markov chain algorithm in which the transition probability
to go from state $i$ to state $j$ with respective energies $E_{i}$
and $E_{j}$ is\begin{equation}
W_{ij}=\min(1,e^{-\Delta\omega(E)}).\label{eq:W_ij}\end{equation}
This choice leads to an asymptotic energy distribution probability
given by 

\begin{equation}
H(E)\propto\exp(S(E)-\omega(E)),\label{p(e)}\end{equation}
where the entropy is defined as the logarithm of the (unknown) density
of energy states, $S(E)=\ln\left(n(E)\right)$. In principle, this
relation allows a calculation of the entropy $S(E)$ (up to an irrelevant
constant $S_{0}$) from a numerical determination of the distribution
$H(E)$ \textsl{for any value of the energy $E$} simply as: \begin{equation}
S(E)=S_{0}+\omega(E)+\ln H(E)\label{SE}\end{equation}
 However, this is not always efficient for any choice of the function
$\omega(E)$. Consider, for example, the widely used Metropolis choice,
$\omega(E)=\beta E$ (with $\beta=1/T$, the inverse temperature).
In many systems, the resulting distribution $H(E)$ has the shape
of a bell curve, usually approximated by a Gaussian distribution (Figure
\ref{esquemas}). Since it is very unlikely to generate statistically
significant configurations in the tails of the distribution for $H(E)$,
the usefulness of the formula (\ref{SE}) is limited to values of
$E$ not too far from the mean value $\mu_{\beta}$. {}``Not too
far\char`\"{} means explicitly that the above formula is limited to
those values of $E$ such that $|E-\mu_{\beta}|<\alpha\sigma_{\beta}$
with $\alpha\sim2-4$. The mean value $\mu_{\beta}$ and the variance
$\sigma_{\beta}^{2}$ of the distribution $H(E)$ satisfy: \begin{equation}
\left.\frac{dS}{dE}\right|_{\mu_{\beta}}=\beta;\hspace{0.5cm}\left.\frac{d^{2}S}{dE^{2}}\right|_{\mu_{\beta}}=-\frac{1}{\sigma_{\beta}^{2}}\end{equation}

A clever choice for $\omega(E)$ can greatly improve the range of
values of $E$ for which the formula (\ref{SE}) is useful. For instance,
if we were to chose $\omega(E)\propto\ln(n(E))$ then the resulting
distribution $H(E)$ would be constant in $E$ and all the energy
values would be sampled with the same frequency. However, it is clear
that this is impossible since $n(E)$ is precisely the function we
want to determine.

Berg's scheme uses a series of functions $\omega_{i}(E)$, each one
of them gives information on $n(E)$ for a range of energy values.
The initial choice $\omega_{0}(E)=0$ (equivalent to a Metropolis
choice at infinite temperature) provides a histogram $H_{0}(E)$ from
which one derives an estimation of the entropy, $S_{0}(E)$, valid
for those values of $E$ visited in a statistically significant way.
After this stage, a new simulation is performed with $\omega_{1}(E)=S_{0}(E)$
in the region visited in the previous simulation, from which we obtain
an entropy estimate $S_{1}(E)$ valid in another range of energies,
and so on. Berg proposes a recursion scheme that allows systematic
corrections of $\omega_{i}(E)$ at all visited energies, and ensures
the convergence to the true entropy for any energy, so ensuring a
flat energy histogram.%

This procedure has some limitations. To begin with, the entropy can
only be estimated inside the energy range visited in the last simulation
and a bad choice for the entropy outside this region can severely
limit the exploration of lower energies; secondly, rarely visited
energies introduce a large error in the estimated entropy (hence the
need for recursion introduced by Berg in order to minimize this error).
Furthermore, the fact that one counts visits in each energy implies
that, for continuous systems, the energy spectrum must be discretized. 

Thermodynamic functions such as the entropy can be treated as continuous
functions of energy, both for discrete and continuous spectra, for
not too small systems. We make use of the Gaussian approximation:
\begin{equation}
S(E)\approx S(\mu_{\beta})+\beta(E-\mu_{\beta})-\frac{1}{2\sigma_{\beta}^{2}}(E-\mu_{\beta})^{2}\end{equation}
 to propose the sequence of weight functions $\omega(E)$.

\begin{figure}
\begin{centering}\includegraphics[width=0.8\columnwidth]{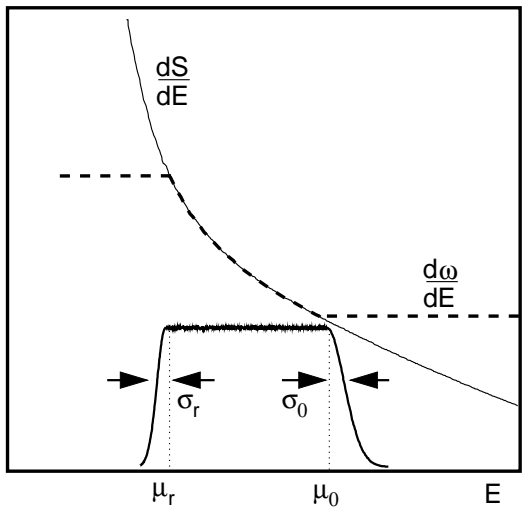}\par\end{centering}

\caption{\label{cap:esquema_cfp}In the current proposal, the weight function
$\omega(E)$ approximates the entropy between two energies $\mu_{r}$
and $\mu_{0}$ and is linear in energy outside this region, $d\omega/dE=\beta_{r}$
for $E<\mu_{r}$ and $d\omega/dE=\beta_{0}$ for $E>\mu_{0}$.}
\end{figure}

After running an initial simulation at a inverse temperature $\beta_{0}$,
with $\omega_{0}=\beta_{0}E$, and measuring $\mu_{0}$ and $\sigma_{0}$,
we modify the weight function to $\omega_{1}(E)$ defined by: 

\begin{equation}
\frac{d\omega_{1}(E)}{dE}=\left\{ \begin{array}{ccc}
\beta_{0} &  & E>\mu_{0}\\
\beta_{0}-\frac{1}{\sigma_{0}^{2}}(E-\mu_{0}) &  & \mu_{1}<E<\mu_{0}\\
\beta_{1} &  & E<\mu_{1}\end{array}\right.\label{w(e) por ramos}\end{equation}
 with $\mu_{1}=\mu_{0}-\alpha\sigma_{0}$ and $\beta_{1}=\beta_{0}+\alpha/\sigma_{0}$.
In a new simulation with transition rates $W_{ij}=\min(1,e^{-\Delta\omega_{1}(E)})$
we obtain $H(E)\simeq\texttt{constant}$, for energies such that $\mu_{1}<E<\mu_{0}$,
while for $E<\mu_{1}$, $H(E)$ is a {}``half Gaussian'' with maximum
at $\mu_{1}$. It is straightforward to estimate $\sigma_{1}$ from
the average of $\left(E-\mu_{1}\right)^{2}$ for $E<\mu_{1}$. We
are therefore be able to add another branch to $\omega(E)$, and so
on, until we reach the lowest temperature we wish to study. On the
iteration of order $r$ the histogram is flat between $\mu_{0}$ and
$\mu_{r}$ and half Gaussian below $\mu_{r}$ (Figure \ref{cap:esquema_cfp}).
Using $\omega(E)=\beta_{0}$ for $E>\mu_{0}$, we effectively restrict
the simulation to energies below $\mu_{0}$, apart from a Gaussian
tail above this energy. 

The fact that $\beta(E)=\texttt{const}$ in the unexplored energy
regions, corresponding to Boltzmann sampling, means that we can use
all techniques developed for the Metropolis algorithm in order to
be confident on the results obtained in this region, before moving
on to lower energies. These regions of canonical sampling can also
be used to restrict the simulation to a specific temperature range,
for instance around a critical temperature, which can dramatically
increase the efficiency and precision of the simulation. The method
can also be refined by keeping higher order terms in the expansion
of the entropy which can be obtained by measuring higher order moments
of the energy. We have successfully used an expansion of $S(E)$ up
to fourth order terms.

\subsection{Comparison with histogram based recursion}

We compared an implementation of this proposal based on classical
fluctuation theory (CFP) with the multicanonical recursion as described
in \cite{Ber03} in a simulation of a two dimensional Ising model:\[
H=-J\sum_{<ij>}s_{i}s_{j}.\]
 Except for the method of estimating the density of states (or rather,
its logarithm, the entropy), the results were obtained using exactly
the same code, specifically with the same number of steps per run.
We chose the number of Monte Carlo steps (MCS) in each iteration (run)
to increase linearly with the iteration number, $r$, specifically
as $r\times10^{5}\,\texttt{MCS}$. This is a rather arbitrary choice
which is necessary for comparison purposes. In fact, our proposal
allows us to set the number of steps in the yet unexplored energy
region as the criteria for moving on to the next run, which, in turn,
allows better statistical control over the next estimate for the entropy.
For a fair comparison between the two algorithms we choose $\beta_{0}=0$
for CFP and impose $\beta(E)=0$ in the histogram based method for
$E>\mu_{0}$. In this way the algorithms only explore the positive
temperature region of the spectra.

\begin{figure}
\begin{centering}\includegraphics[width=0.9\columnwidth,keepaspectratio]{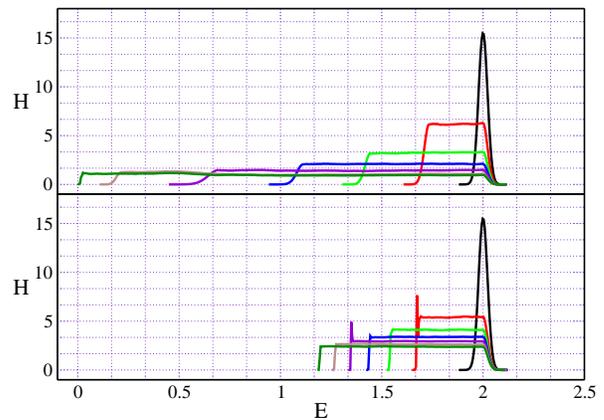}\par\end{centering}

\caption{\label{cap:histogramas}A comparison of the evolution of the multicanonical
recursion in the present scheme (upper panel) and the one described
in \cite{Ber03} (lower panel). The simulation was done for a Ising
square sample with $L=64$, and, for simplicity, we only show histograms
from 1 out of each 5 runs.}
\end{figure}

\begin{figure}
\begin{centering}\includegraphics[width=0.8\columnwidth,keepaspectratio]{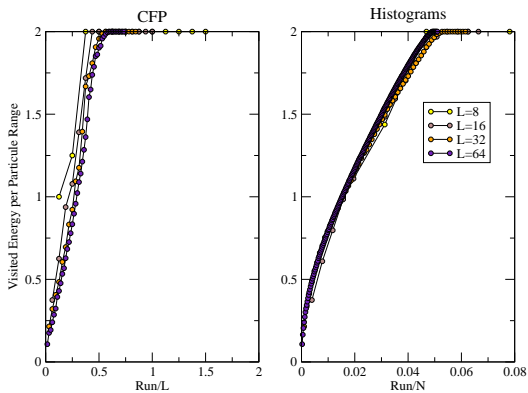}\par\end{centering}

\caption{\label{cap:range}Full range of visited energies ($E_{r}-E_{0})/N$,
as a function of the system size and number of iterations, $r$, for
current method (left panel) and for histogram based recursion (right
panel); in our moment based method $r$ scales $L=\sqrt{N}$ whereas,
with histograms, it scales with $N$.}
\end{figure}

Figure \ref{cap:histogramas} shows 7 histograms out of 35 runs of
simulations on a $64\times64$ Ising model, with $r\times10^{5}\texttt{\, MCS}$
per run, which was enough to reach the ground state with our algorithm.
Several more runs are required using an histogram based method (lower
panel of fig.~\ref{cap:histogramas}). It is important to note that
the full range of visited energies, $\sim\left(\mu_{0}-\mu_{r}\right)$,
does not increase linearly with run number, $r,$ because the spectral
range added in each run shrinks as one approaches the ground state.
In this respect the difference in performance of the two methods is
quite remarkable. Histogram based methods do not provide an estimate
for the entropy outside the previously visited energy range. As a
result, for energies in the previously unvisited range, $H(E)$ decays
as\[
H(E)\sim\exp\left[S(E)\right]\sim\exp\left[-\beta_{r}(E_{r}-E)\right]\]
where $E_{r}$ is the lowest energy of the previous run and $\beta_{r}=\beta(E_{r}).$
Our method uses $\omega(E)=\beta_{r}E$ which leaves \[
H(E)\sim\exp\left[S(E)-\omega(E)\right]\sim\exp\left[-\frac{1}{2\sigma_{r}^{2}}(E_{r}-E)^{2}\right]\]
 Since $\sigma_{r}^{2}$ scales linearly with $N$, the added energy
range in each iteration scales differently in the two methods. In
terms of energy per particle, the added energy range per run, $\Delta\epsilon=\Delta E/N$,
scales as $1/L=N^{-1/2}$ in the current proposal instead of $\left(\beta_{r}N\right)^{-1}$
in histogram based methods. In Figure~\ref{cap:range} we plot the
visited energy per particle range after $r$ iterations, for various
system sizes, as a function of $r/L$ for the CFP method and $r/N$
for histogram based method. The collapse of the curves for the various
systems sizes shows that the number of runs required to cover the
same energy per particle range scales with with $L=\sqrt{N}$ in the
current proposal (left panel) and $N$ in histogram based methods
(right panel). 

The two methods perform similarly for small systems, but the advantage
of CFP method becomes obvious for large $N$ and no amount of fine
tuning can disguise it. 

Nevertheless, other alternative schemes to Berg's recursion have already
been proposed like Wang-Landau sampling \cite{WL01b-a} or the transition
matrix method \cite{WTS99}. As will be seen shortly, the main advantage
of our method is that, unlike these previous methods, it produces
an analytic approximation to the microcanonical inverse temperature,
$\beta(E)$, in an increasing energy range, right from the start of
the simulation. That proves an asset in the implementation of procedures
designed to overcome the slow down with system size that affects multicanonical
simulations \cite{DTW+04,THT04,WKC+05}.

\section{Optimization of tunneling times}

In a recent publication, Trebst, Huse and Troyer showed that it is
possible to decrease significantly the tunneling times of a multicanonical
simulation, by abandoning the requirement of a flat histogram \cite{THT04}.
Our procedure of construction of the statistical entropy is well suited
to implement their optimization method, right from the start of the
simulation, without having to construct an approximation to $n(E)$
for the entire spectrum.

The procedure proposed in \cite{THT04} minimizes the average time
required to span the gap between two fixed energies in the spectrum,
$E_{-}$ and $E_{+}$.%
\begin{figure}
\begin{centering}\includegraphics[width=0.8\columnwidth]{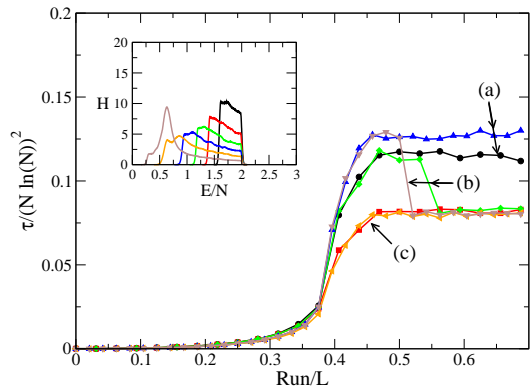}\par\end{centering}

\caption{\label{cap:CPP_Opt}Tunneling time (scaled by $(N\log N))^{2}$,
for system sizes $N=48\times48$ and $N=64\times64$, as an function
of the recursion run divided by $L$ which is proportional to the
energy per particle range. (a) CFP without optimization; (b) for an
application of the optimization procedure only after an estimation
of density of states in the chosen range of temperatures; (c) our
implementation of the optimization. The inset shows the histograms
obtained for the sample with $L=64$ in several runs during the exploration
of the energy spectrum with optimization.}
\end{figure}
 To achieve this purpose, one must distinguish each energy entry during
the simulation according to which of the two energies, $E_{-}$ or
$E_{+}$, was visited last. We can thus measure separately $n_{-}(E)$,
the number of visits to energy $E$ occurring when the simulation
has visited $E_{-}$ more recently than $E_{+}$, and $n_{+}(E)$
for the other way around. To minimize the tunneling times between
the two energies $E_{-}$ and $E_{+}$ one must choose an asymptotic
energy distribution that satisfies\begin{equation}
H(E)\propto\frac{df(E)}{dE}\label{eq:criterio_opt}\end{equation}
where $f(E)=n_{+}(E)/H(E)$. The implementation proposed in \cite{THT04}
used the knowledge of the density of states, $n(E)$ in the whole
spectrum to measure the $f(E)$, following with a recursion procedure
that converges to the asymptotic distribution which satisfies eq.~\ref{eq:criterio_opt}.
In our method of construction of the density of states there are,
at each step, two energies which are the current boundaries of the
known density states, namely $\mu_{0}$ (that remains fixed) and $\mu_{r}$
(that changes with each run, $r$). Therefore, by using $E_{-}=\mu_{r}$
and $E_{+}=\mu_{0}$ we can measure $f(E)$. In the first run, for
$\mu_{1}<E_{i}<\mu_{0}$, we observe $H_{1}(E_{i})\propto\exp\left(-\omega_{1}(E)\right)$,
with $\omega_{1}(E)$ defined in eq.~\ref{w(e) por ramos}, rather
than the optimal distribution $H^{opt}(E)\propto\exp\left[\ln\left(df_{1}/dE\right)\right]$.
To converge to the optimal distribution (changing the weight alters
$f(E)$), we use, in the next run, the geometric mean $\sqrt{H_{1}H^{opt}}$
in the interval $\mu_{1}\le E\le\mu_{0}$ \cite{THT04}: the weight
factor becomes $-\omega_{2}(E)+\phi_{2}(E)$, where\[
\phi_{2}(E)=\frac{1}{2}\ln\left(\frac{df_{1}}{dE}\right)\qquad\mbox{for }\mu_{1}\le E\le\mu_{0},\]
with constant values of $\phi_{2}(\mu_{1})$ for $E<\mu_{1}$ and
$\phi_{2}(\mu_{0})$ for $E>\mu_{0}$. Notice that the correction
to the microcanonical temperature, $\beta(E)$, is $-d\phi_{2}(E)/dE$
which is zero for $E<\mu_{1}$. This choice ensures the convergence
to the criterion of eq.~\ref{eq:criterio_opt} in the range where
the entropy is already known and where $f(E)$ was measured, $\mu_{1}<E<\mu_{0}$,
and gives a flat histogram in the region where the entropy was estimated
by calculating moments of the Gaussian tail, i.e. $\mu_{2}<E<\mu_{1}$.
This procedure is iterated in the following runs with $\phi_{r}(E)$
defined as, \[
\phi_{r}(E)=\frac{1}{2}\left[\phi_{r-1}+\ln\left(\frac{df_{r-1}}{dE}\right)\right];\qquad\mu_{r-1}\le E\le\mu_{0}\]
with constant values $\phi_{r}(\mu_{r-1})$ for $E<\mu_{r-1}$ and
$\phi_{r}(\mu_{0})$ for $E>\mu_{0}$. To extract the numerical derivative
of $f(E)$, avoiding the difficulties of the fluctuations in histogram
entries, we use the natural scale afforded by $\sigma_{\beta}$, and
calculate $df/E$ as\[
\left.\frac{df}{dE}\right|_{\mu_{\beta}}=\frac{\left\langle f(E)\left(E-\mu_{\beta}\right)\right\rangle _{\beta}}{\sigma_{\beta}^{2}}.\]

In figure \ref{cap:CPP_Opt} we plot the average tunneling time, divided
by $(N\ln N)^{2}$, for two different system sizes, as an function
of the run number of the multicanonical recursion divided by $L=\sqrt{N}$;
simulations with the same horizontal coordinate correspond to the
same energy per particle range. Curves (a) correspond to the situation
without optimization, and the average tunneling times vary faster
than $(N\ln N)^{2}$; the two sizes do not collapse to a single curve.
In case (c), with the optimization carried out while the density of
states is being determined, the curves for the two system sizes track
each other. If the optimization correction is only performed after
full exploration of the energy spectrum, curve (b), there is no further
gain in tunneling time, as the average tunneling times of (b) merge
with curve (c). This observation clearly supports our suggestion that
optimization can be implemented while the density of states is being
constructed. In this fashion, it not only reduces tunneling times
when $n(E)$ is known, but also speeds up the actual calculation of
$n(E)$. As found in \cite{THT04}, the inset shows a strong signature
of the optimizing procedure in the critical energy where the diffusivity
is low.

\section{Cluster Dynamics}

An alternative way to improve the efficiency of multicanonical simulations
consists in changing the specific dynamics of the Markov chain, \emph{i.e.,}
the algorithms used to propose and accept configuration moves.

In canonical ensemble simulations, cluster update algorithms like
Wolff's \cite{WOL89}, Swendsen-Wang \cite{SW87} or Niedermeyer's
\cite{NIE88} have proved very effective in overcoming critical slowing
down of correlations. Several proposals have been presented to generalize
cluster update approaches in multicanonical ensemble simulations,
either using spin-bond representations of the partition function,
\cite{JK95c,JK95d,YK02,WKC+05}, or cluster building algorithms based
on alternative ways of computing the microcanonical temperature, $\beta(E)$
\cite{RD05b,RD05}.

Wolff's cluster algorithm \cite{WOL89} provides a clever way of growing
a cluster of parallel spins which can be flipped with probability
1, and still maintain the required Boltzmann asymptotic distribution.
This remarkable possibility is intimately related to the fact that
$\omega(E)$ is linear in energy, $\omega(E)=\beta E$, in a Metropolis
simulation.

In a multicanonical simulation, each step of the corresponding Markov
chain occurs with the same probability as that of a canonical ensemble
simulation, with an effective temperature $\beta_{i}$ chosen as $\beta_{i}=\beta(E_{i})$,
$E_{i}$ being the energy of the current configuration; hence the
designation {}``Multicanonical''. With this in mind, the simplest
way of implementing cluster dynamics in a multicanonical simulation
is to use $\beta_{i}=\beta(E_{i})$ to grow a cluster exactly as proposed
in Wolff's algorithm. However, since the reverse path implies a different
value of $\beta$, \textbf{$\beta_{j}=\beta(E_{j})$}, where $E_{j}$
is the energy of the next configuration in the chain, we must include
an acceptance probability to ensure detailed balance.

\begin{figure}
\begin{centering}\includegraphics[width=0.8\columnwidth]{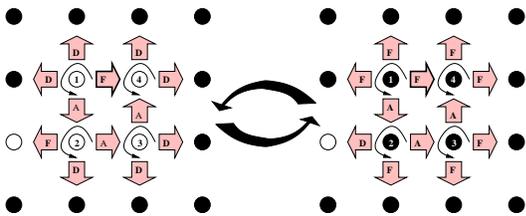}\par\end{centering}

\caption{\label{cap:Esquema_cluster}This scheme shows, on the left, how,
starting from a given spin (marked as $1$), the bonds to neighbouring
spins are inspected and marked according to the algorithm. On the
right, are shown the resulting configuration and the way the initial
state can be reached from it. }
\end{figure}

In figure~\ref{cap:Esquema_cluster} we illustrate a move involving
the flipping of four spins (labelled $1$ to $4$) on the left, and
the reverse move on the right. The site marked with the number 1 has
been chosen with uniform probability. If a bond connects spin 1 to
a neighboring spin \emph{parallel to} 1 it is added to the cluster
with a probability $p_{i}$ and rejected with probability $1-p_{i}$.
This step is then repeated for the neighbours of the initial spin
which were added to the cluster, until the process stops and there
are no further bonds that can be aggregated to the cluster. The probability
of generating a cluster with $n_{a}$ accepted bonds, in which $n_{r}$
bonds to spins parallel to the initial one were inspected and rejected,
is given by:\[
G_{i\to f}=p_{i}^{n_{a}}(1-p_{i})^{n_{r}}.\]
It is important to note that $n_{r}$ includes a number of rejected
bonds that now link spins inside the cluster (like the bond from spin
$1$ to spin $4$ in fig.~\ref{cap:Esquema_cluster}). We write $n_{r}=n_{p}+n_{f}$,
where $n_{p}$ counts the number of such bonds and $n_{f}$ is the
number of bonds from spins in the cluster to parallel spins outside
the cluster. This distinction is important when considering the reverse
move. 

The difference in energy between the final and initial configuration
is determined only by the frontier of the cluster; it is given as
$\Delta E=-2J(n_{d}-n_{f})$ where $n_{d}$ is the number of bonds
to spins opposite to the spins in the cluster. Let us now consider
the reverse move, which requires us to select the same cluster in
the same order with the spins now reversed we respect to the original
state.

Referring once again to figure \ref{cap:Esquema_cluster}, one can
see that the $n_{a}$ bonds that were accepted in the direct move
(left) with probability $p_{i}$ must be accepted in the reverse move
(right), each with probability $p_{j}$; the $n_{d}$ bonds to opposite
spins in the direct move now connect to spins parallel to those in
the cluster and must be rejected probability $(1-p_{j})$; the bonds
to spins that were rejected in the direct move ($n_{f}$) are, in
the reverse process, bonds to opposite spins and are, hence, rejected
with probability $1$; finally the $n_{p}$ bonds that were rejected
but link spins inside the cluster, must now also be rejected. In other
words, for the direct move

\begin{equation}
G_{i\to f}=p_{i}^{n_{a}}(1-p_{i})^{n_{f}}(1-p_{i})^{n_{p}}\end{equation}
while, for the reverse process, \begin{equation}
G_{f\to i}=p_{j}^{n_{a}}(1-p_{j})^{n_{d}}(1-p_{j})^{n_{p}}.\end{equation}
Wolff's algorithm corresponds to choosing $p_{i}=p_{j}=1-\exp(-2J\beta)$
which implies that \[
\frac{G_{f\to i}}{G_{i\to f}}=(1-p_{i})^{n_{d}-n_{f}}=e^{\beta\Delta E}.\]
The detailed balance condition for Boltzmann's equilibrium distribution
is obtained for an acceptance probability of 1 for flipping the cluster.
To ensure an asymptotic distribution proportional to $\exp\left[-S\left(E\right)\right],$
the detailed balance requires an acceptance probability given by\begin{equation}
A_{i\to f}=\min\left(1,\frac{G_{f\to i}}{G_{i\to f}}e^{-\Delta S(E)}\right).\end{equation}

\begin{figure}
\begin{centering}\includegraphics[width=0.8\columnwidth]{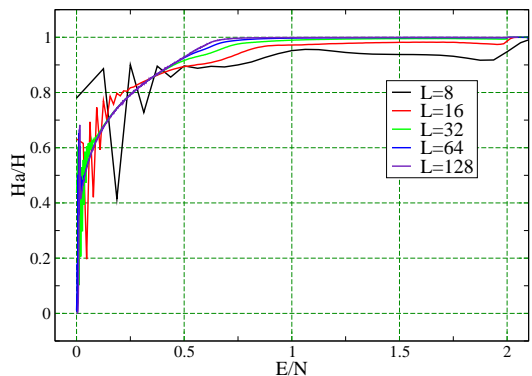}\par\end{centering}

\caption{\label{cap:aceitacao_cluster}Acceptance ratio of the Wolff 's cluster
algorithm, with a microcanonical $\beta(E)$, for the 2D ferromagnetic
Ising model. In the paramagnetic phase the acceptance ratio grows
to $1$ with the increase of system size; in the ferromagnetic phase
the acceptance ratio tends to a finite value, smaller than $1$, as
$N$ grows.}
\end{figure}

If we choose $p_{i}=1-e^{-2\beta_{i}J}$ where $\beta_{i}=\beta(E_{i})$,
we find that this acceptance probability remains close to $1$ for
most of the energy range (see fig.~ \ref{cap:aceitacao_cluster}),
falling only at very low temperatures. This behaviour is in strong
contrast with the one for single spin flip dynamics, where the acceptance
rate is only $1$ at the maximum of the density of states. If the
inverse temperatures of the initial and final states are close, $\beta_{i}\approx\beta_{j}$,
the acceptance rate becomes\begin{eqnarray}
A_{i\to f} & \approx & \min\left(1,\left(1-p_{i}\right)^{n_{f}-n_{d}}e^{-\Delta S(E)}\right)\\
 & = & \min\left(1,e^{\beta_{i}\Delta E}e^{-\Delta S(E)}\right)\end{eqnarray}
where we used $\Delta E=-2J(n_{d}-n_{f})$ to obtain the second expression.
For small energy differences, $\Delta S\approx\beta_{i}\Delta E$
and the acceptance rate becomes close to unity. 

\begin{figure}
\begin{centering}\includegraphics[width=0.8\columnwidth]{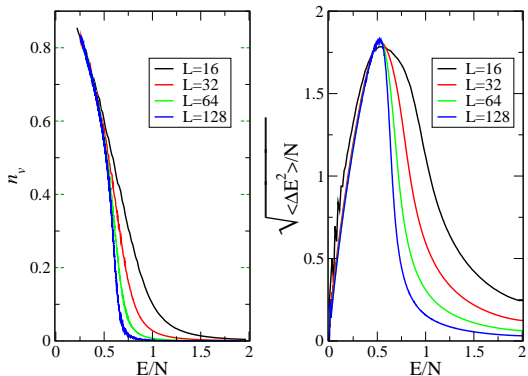}\par\end{centering}

\caption{\label{cap:deltaE_cluster}Two regime behavior of the Wolff's cluster
algorithm for the 2D ferromagnetic Ising model, with a microcanonical
temperature $\beta(E)$. The left panel represents the equal energy
average of the fraction of visited spins, $n_{v}$, during the cluster
growing. In the ferromagnetic phase the number of visited spins is
of the order of $O(N)$ and in the paramagnetic phase is of $O(1)$.
The right panel it is represents the equal energy average of the excitations,
$\Delta E$. In the paramagnetic phase $\sqrt{\left\langle \Delta E^{2}\right\rangle }$
becomes independent of the system size, while in the ferromagnetic
phase it scales as $\sqrt{N}$.}
\end{figure}

In the case of the 2D ferromagnetic Ising model, the average excitation
energy, $\Delta E$, is of $O(\sqrt{N})$ for energies below $\mu_{c}$
($\mu_{c}=\mu_{\beta_{c}}$ where $\beta_{c}$ is the inverse critical
temperature), of $O(1)$ for energies above $\mu_{c}$, with a crossover
between these regimes in the neighborhood of $\mu_{c}$ (fig.~\ref{cap:deltaE_cluster}).
This behaviour reflects a huge difference with respect to single spin
flip dynamics (SSF), where $\Delta E$ is always of $O(1)$. A similar
behaviour exists for the number of spins, $n_{v}$ that are inspected
in each call of the Markov chain: $n_{v}\sim O(1)$ for $E>\mu_{c}$
and $n_{v}\sim O(N)$ for $E<\mu_{c}$.%
This two regime behaviour introduces an additional complexity in this
method. In particular, the tunneling time measured in Markov chains
calls, no longer scales as the computational time with system size,
since Markov chain calls can take a computational time of order $O(N)$.
We therefore redefine the time scale so that a Markov chain call in
a state of energy $E$ corresponds to a time span of $n_{v}(E)$,
the average number of inspected spins in the cluster buildup process.

We now consider the system's coarse grained random walk in energy
space. When the energy is close to $E$, the mean square energy change
in $M$ Markov chain calls is \[
\left\langle \Delta E^{2}\right\rangle _{M}=\left\langle \Delta E^{2}\right\rangle \times M\]
and this occurs in a time $\tau_{M}=M\times n_{v}(E)$. With time
measured in this way, the diffusivity of this random walk in energy
space is \begin{equation}
D(E)\propto\frac{\left\langle \Delta E^{2}\right\rangle }{n_{v}(E)}.\label{eq:D_E_n_v}\end{equation}
 On the other hand, the probability that the system is at energy $E$
is proportional to \begin{equation}
H_{v}(E)=H(E)\times n_{v}(E),\label{eq:hv_E}\end{equation}
 since each visit to energy $E$ lasts a time $n_{v}(E)$. The probability
current is given, quite generally, by\[
j=\rho(E)V(E)-\frac{d}{dE}\left(D(E)\rho(E)\right)\]
where, in equilibrium, $j=0$, and \[
\rho(E)=\rho_{0}(E)\equiv\frac{H_{v}(E)}{\int dEH_{v}(E)}.\]
$V(E)$ is a bias field, in general non-zero, which, together with
$D(E)$, determines the equilibrium distribution.

\begin{figure}
\begin{centering}\includegraphics[width=1\columnwidth]{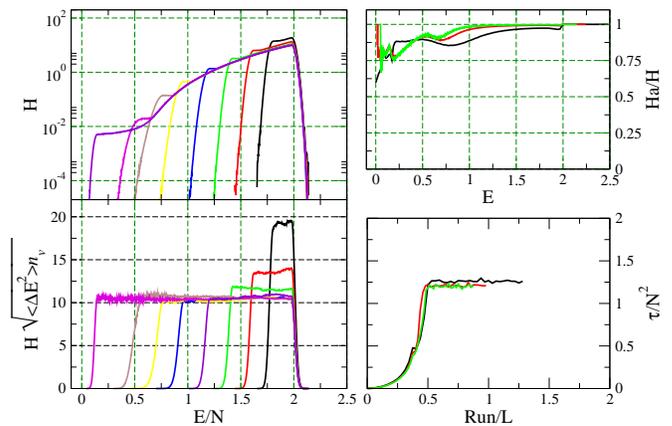}\par\end{centering}

\caption{\label{cap:Cluster_CE}Results of the CFP recursion to construct
an histogram given by eq. \ref{eq:h_e_comp_eff}. (a) Histogram of
the simulation for several steps of the CFP recursion; (b) Acceptance
ratio; (c) ($H(E)\times\sqrt{n_{v}(E)\left\langle \Delta E^{2}\right\rangle }$)
; (d) Tunneling time (scaled by $N^{2}$) versus the run of the recursion
(scaled by $L$).}
\end{figure}

It can be shown \cite{JJJ,THT04} that the tunneling times of a random
walk with a given diffusivity, $D(E)$, can be minimized with an optimal
choice of bias field $V(E)$. The corresponding equilibrium distribution
is given by

\[
\rho_{0}(E)=\frac{1}{\sqrt{D(E)}}.\]
Using equations~\ref{eq:D_E_n_v} and \ref{eq:hv_E}, this condition
becomes\begin{equation}
H(E)\propto\frac{1}{\sqrt{\left\langle \Delta E^{2}\right\rangle n_{v}(E)}}\label{eq:h_e_comp_eff}\end{equation}
This variation of $\log\left(H(E)\right)$, relative to a flat histogram,
is of order $O(\log(N))$, and, therefore, histograms remain broad,
covering the entire spectrum, but are no longer flat as shown in panel
(a) of figure \ref{cap:Cluster_CE}. On panel (d) it is shown that
the tunneling time for this simulation scales has $N^{2}$ as expected
from a simple diffusion.%

The optimization procedure presented in reference \cite{THT04}, in
the context of N-fold Way dynamics, is closely related to this one
(but not identical) and leads to a choice\[
H(E)\propto\frac{1}{n_{v}(E)}\frac{df}{dE}\]
where $f(E)$ was defined above. We find only a marginal improvement
in tunneling times with respect to the case of a histogram defined
by eq.~\ref{eq:h_e_comp_eff}. These two procedures will be compared
in another publication \cite{JJJ}.

\section{Conclusion}

We have proposed a new method to build the density of states in an
multicanonical simulation. The method is based on the calculation
of moments of the energy distribution. It avoids the use of histograms
and can just as easily be implemented for continuous as for discrete
systems. It leads to a piecewise analytic approximation to the microcanonical
inverse temperature $\beta(E)$. In any stage of the simulation there
are two well defined energies, $E_{-}$ and $E_{+}$, that limit the
range in which $\beta(E)$ is known. Therefore the method can be applied
without difficulty to a predefined temperature range such as a neighborhood
of a critical temperature. %
\begin{figure}
\begin{centering}\includegraphics[width=0.8\columnwidth]{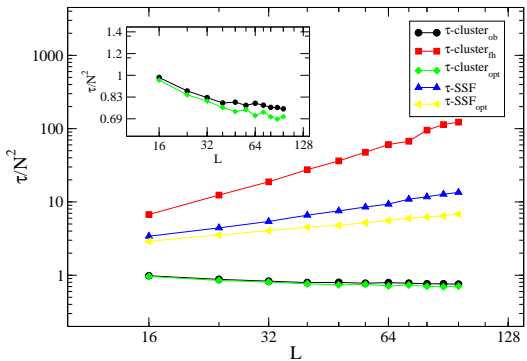}\par\end{centering}

\caption{\label{cap:Tt}Scaling of tunneling time with system size in various
broad histogram methods: (SSF) straight forward multicanonical simulation
with single spin flip dynamics; ( SSF$_{opt}$) optimized single spin
flip dynamics; $(\textrm{cluster}_{ob}$) optimized bias for the measured
$D(E)$; ($\textrm{cluster}_{opt}$) optimized ensemble with cluster
dynamics; ($\textrm{cluster}_{fh}$) flat histogram with cluster dynamics. }
\end{figure}

We have also demonstrated the usefulness of this method in the implementation
of various optimization schemes that render the simulation more efficient.
In figure~\ref{cap:Tt} we sum up the results we obtained for the
scaling of tunneling times with the system size. In general the scaling
of the average tunneling time is $\tau\sim N^{2+z}$ . In a straightforward
multicanonical simulation with a SSF dynamics, $z=0.39$. Using the
optimization procedure of \cite{THT04} we confirm that $\tau\sim(N\ln N)^{2}$.
For a generalization of Wolff's cluster method for the multicanonical
ensemble we found a biased random walk in energy with $z=0.82$. We
proposed a new method for reducing tunneling time of cluster update
simulations which adjusts the bias of the random walk in energy space.
In this case, ($\textrm{cluster}_{ob}$) and also for optimized ensemble
simulation with cluster algorithm, proposed in \cite{THT04} ($\textrm{cluster}_{opt}$),
the results are compatible with $z=0$. In the ($\textrm{cluster}_{ob}$)
method, however, one avoids the necessity to calculate of the derivative
of $f(E)$, required for the ($\textrm{cluster}_{opt}$) method of
\cite{THT04}. In terms of \emph{actual computer time,} we also found
that the amplitudes of the scaling laws are considerably smaller for
the optimized cluster methods, than for SSF dynamics.

The authors would like to thank J. Penedones for very helpful discussions.
This work was supported by FCT (Portugal) and the European Union,
through POCTI (QCA III). Two of the authors, JVL and MDC, were supported
by FCT grants numbers SFRH/BD/1261/2000 and SFRH/BD/7003/2001, respectively.

\bibliographystyle{apsrev}
\bibliography{Multicanonical}

\end{document}